\documentclass[12pt,halfline,a4paper]{ouparticle}
\usepackage{natbib}
\usepackage[official]{eurosym}
\usepackage{hyperref}
\hypersetup{
colorlinks = true,
linkcolor = cyan,
citecolor = magenta,
urlcolor = blue,
}
\begin{document}

\title{Environmental Applications of the Coase Theorem}

\author{Tatyana Deryugina\thanks{University of Illinois at Urbana-Champaign, Department of Finance; NBER; IZA; e2E. \href{mailto:deryugin@illinois.edu}{\texttt{deryugin@illinois.edu}}}
\and
Frances Moore\thanks{University of California Davis, Department of Environmental Science and Policy, Davis, CA,\\
95616, USA.
\href{mailto:fmoore@ucdavis.edu}{\texttt{fmoore@ucdavis.edu}}}
\and
Richard S.J. Tol\thanks{University of Sussex, Department of Economics, Falmer, BN1 9SL, United Kingdom; Institute for Environmental Studies, Vrije Universiteit, Amsterdam, The Netherlands; Department of Spatial Economics, Vrije Universiteit, Amsterdam, The Netherlands; Tinbergen Institute, Amsterdam, The Netherlands; CESifo, Munich, Germany; Payne Institute for Earth Resources, Colorado School of Mines, Golden, CO, USA. \href{mailto:r.tol@sussex.ac.uk}{\texttt{r.tol@sussex.ac.uk}}}
}

\abstract{The Coase Theorem has a central place in the theory of environmental economics and regulation. But its applicability for solving real-world externality problems remains debated. In this paper, we first place this seminal contribution in its historical context. We then survey the experimental literature that has tested the importance of the many, often tacit assumptions in the Coase Theorem in the laboratory. We discuss a selection of applications of the Coase Theorem to actual environmental problems, distinguishing between situations in which the polluter or the pollutee pays. We find that most substantive examples of Coase-like bargaining involve more than two parties, though whether the outcomes of these bargains were Pareto optimal rather than merely Pareto improving is not clear. While limited in scope, Coasian bargaining over externalities offers a pragmatic solution to problems that are difficult to solve in any other way.}

\date{\today}


\maketitle

\section{Introduction}
\label{sc:intro}
The Coase Theorem \citep{Coase1960} is a central result in economics.\footnote{Coase also made key contributions to the theory of the firm \citep{Coase1937} and the behaviour of a durable-goods monopolist \citep{Coase1972}.} It shows how, under certain conditions, economic actors can arrive at an efficient solution to an externality \textit{without direct government involvement}. Prior to Coase's seminal paper, economists thought that externalities, which are at the heart of environmental economics, necessitate government regulation, particularly taxation \citep{Pigou1920}. Since then, the Coase result has sometimes been used to argue that environmental externalities do not necessitate government regulation beyond the establishment and enforcement of property rights. Skepticism remains, however, regarding the applicability of Coase's theoretical result to real-world environmental problems.

Coase's aim was to ``expos[e] the weaknesses of Pigou's analysis" \citep{Coase1991}. We study to what degree he has, not in theory but in practice. \footnote{\citet{Medema2020} gives an excellent overview of how the Coase Theorem came about, the initial response by the economic profession, and its eventual impact. \citet{Lai2011} covers similar ground. Lai and Medema, however, do not provide much detail on practical applications in environmental policy, the focus of this paper.}
Specifically, we evaluate the extent to which the Coase Theorem has been or could be used to solve actual environmental problems as opposed to being a theoretical curiosity, drawing on experimental evidence and real-world examples.

All major textbooks in environmental economics discuss the Coase Theorem at greater or lesser length. The two central tenets of the theorem are always there:\footnote{\citet{Laurent2020} is the exception, equating the Coase Theorem to tradable emission permits.} (1) a Pareto optimal outcome can be obtained through bargaining if initial property rights are clearly assigned, and (2) that outcome is independent of who initially holds those rights. The textbooks diverge, however, in their assessment of the applicability of the Coase Theorem to actual environmental problems. Indeed, in their review, \citet{CHERRY2013} argue that ''a consensus has not been reached over the validity and importance of the Coase Theorem and how it can be effectively applied to [environmental] policy.'' Six textbooks present the Coase Theorem as an intellectual curiosity with little value in real life \citep{Anderson2019,Hodge1995,Pearce1990,Perman2011,Tietenberg2018,Turner1994}.\footnote{\citet{Endres2012} and \citet{Lewis2019} only mention the Coase Theorem in passing.} \citet{Harris2018} additionally argue that seemingly voluntary transactions may in fact be coercive and thus unjust. In contrast, eight other textbooks emphasize that the Coase Theorem can be used in certain circumstances, while also highlighting the restrictive assumptions of the Coase Theorem and its limited applicability \citep{Baumol1988, Field2009, Goodstein2005,Keohane2016,Kolstad2011,Phaneuf2017,Wills1997,Kahn2020}. Three textbooks go further, presenting the allocation of property rights followed by bargaining as a viable policy option \citep{Berck2011,Hanley2007,Hanley2013}. Environmental economists thus appear to be divided about the practical relevance of the Coase Theorem under realistic conditions that characterize many environmental problems, including information constraints and transaction costs.

To the best of our knowledge, we are the first to review evidence for the Coase Theorem being applied to actual environmental problems. Although there are papers that describe applications of the Coase Theorem to a single case \citep{Hanley1995, Ruml2005, Pirard2012, Folefack2014, Byun2015}, it is difficult to take general lessons from single-case studies. We discuss a number of cases but restrict ourselves to qualitative analysis. Applications of the Coase Theorem are hard to find, potentially because no direct government involvement is required. Documentation may therefore be missing, unless the case is newsworthy or amusing\textemdash such a noisy chemical plant offering chocolates to nearby residents\footnote{See E. Schmidt,  \href{https://www.limburger.nl/cnt/dmf20191220_00138200/sabic-zegt-sorry-met-doosje-bonbons}{Sabic apologizes with chocolates}, De Limburger, 20 Dec 2019.} or the obese man paying the man sitting next to him on a plane \$150 for being allowed to take up some of the latter's seat.\footnote{See R. Hosie, \href{https://www.insider.com/reddit-thread-man-charges-obese-passenger-on-plane-150-for-encroaching-on-seat-2019-3}{A man charged an obese passenger sitting next to him on a plane \$150 for encroaching on his seat, and he was happy to pay it}, Insider, 4 Mar 2019.} Personal anecdotes in which bargaining is used to address externalities abound: A colleague's father offered to buy a rug for his upstairs neighbor to muffle the sound of the neighbor's grandchild dribbling a basketball. A colleague offered brownies and money, and an author offered pears to their respective neighbors if they would quit calling the cops on their practising college bands for noise violations. Bargains like these are not systematically recorded, and so we cannot know whether these are the exceptions or the rule. They are also arguably of less relevance to those who are interested in whether Coasian bargaining can substitute for Pigovian regulation, as Pigovian regulation would not be feasible in such settings due to the difficulty of monitoring, let alone taxing the to-be-regulated entities.

One challenge with surveying real-world applications of the Coase Theorem is that there is non-trivial disagreement as to what constitutes Coasian bargaining. What if, for example, one of the bargaining parties is a local or a national government? Should compensation that takes place \textit{after} the damage has occurred be counted? Strictly speaking, the Coase Theorem applies to bargains between two agents, but its proponents have suggested bargaining as a solution to problems involving larger numbers of actors. Resolving such disagreements is beyond the scope of this article. Therefore, we err on the side of being inclusive and leave it to the reader to decide whether a particular example meets her preferred definition of Coasian bargaining.

A consistent theme that emerges is that many cases where bargaining looks feasible \textit{ex ante} end up in court nonetheless. Cases that settle could be viewed as Coasian bargaining with some transaction costs where the involvement of the legal system helps render the contract between parties enforceable. However, a failure to come to a resolution before a (costly) trial is more consistent with imperfectly specified property rights. Indeed, a review of prominent court cases suggests that disagreement over whether the polluter has the right to inflict the harm in question (in other words, disagreement over the nature of the relevant property rights) is a major reason why more Coasian bargains are not being struck in reality and instead the relevant parties end up in courts. In other situations, there appears to be substantial uncertainty over whether or not a property right will be enforced by the courts. Both the definition of property rights and the enforcement of contracts (including any Coasian bargains negotiated between parties) point to the critical role of public institutions: even if the government itself does not act as a Pigovian regulator, it still plays a central role in creating and maintaining the conditions that enable successful Coasian bargaining.

Strictly speaking, the Coase Theorem is beyond empirical analysis. We observe neither the Pareto optimum nor the counterfactual in which the other party holds the property rights. We therefore necessarily restrict our attention to the observation of people and organizations negotiating over externalities. By the revealed preference argument, if Coasian bargaining is successful, it must have produced a Pareto improvement. How close or far this improvement is from the optimum is not readily observable.

Some recent literature on the Coase Theorem distinguishes between ``entitlements'' and ``property rights'', where ``entitlements'' refers to the initial allocation and ``property rights'' materialize only after negotiations have taken place \citep[e.g.][]{Slaev2017,Slaev2020}. While such a distinction is important in some cases, in this paper we use the term ``property rights'' to refer to the initial delineation of the parties' rights, as this is the more common usage in economics.

The paper proceeds as follows. Section \ref{sc:coase} discusses the Coase Theorem, its key assumption, and influential interpretations. Section \ref{sc:lab} reviews experimental tests of the Coase Theorem. Section \ref{sc:wild} assesses natural evidence, distinguishing between cases where the polluter pays and where the pollutee pays. Section \ref{sc:conclude} charts further research and concludes. 

\section{The Coase Theorem}
\label{sc:coase}

\subsection{Coase in context}
The Coase Theorem was published in the \textit{Journal of Law and Economics}. In its original form, it is not a theorem in the conventional sense of the word. Coase did not formalize his theorem, let alone prove it. There is not a single equation or rigorous definition in the paper.\footnote{According to \citet{Medema2014}, \citet{Stigler1966} coined the term ``Coase Theorem''. Stigler did not, however, restate Coase' insight as a theorem.} Instead, Coase offers a detailed discussion of common law on liability and nuisance. Coase agrees with Pigou that externalities are a problem, but disagrees with Pigou's solution.

\citet[II.XI.11]{Pigou1920} famously developed the theory of externalities, formally defined as situations where the marginal private cost differs from the marginal social cost. He  described a program of taxes or subsidies, which now bears his name, that could internalize these externalities:
\begin{quote}
    ``It follows that, under conditions of simple competition, for every industry in which the value of the marginal social net product is greater than that of the marginal private net product, there will be certain rates of bounty, the granting of which by the State would modify output in such a way as to make the value of the marginal social net product there more nearly equal to the value of the marginal social net product of resources in general, thus\textemdash provided that the funds for the bounty can be raised by a mere transfer that does not inflict any indirect injury on production\textemdash increasing the size of the national dividend and the sum of economic welfare; and there will be one rate of bounty, the granting of which would have the optimum effect in this respect. In like manner, for every industry in which the value of the marginal social net product is less than that of the marginal private net product, there will be certain rates of tax, the imposition of which by the State would increase the size of the national dividend and increase economic welfare; and one rate of tax, which would have the optimum effect in this respect. These conclusions, taken in conjunction with what has been said in the preceding paragraphs, create a presumption in favour of State bounties to industries in which conditions of decreasing supply price \textit{simpliciter} are operating, and of State taxes upon industries in which conditions of increasing supply price from the standpoint of the community are operating.''
\end{quote}
Thus, Pigou argues for the State to intervene to internalize externalities, by imposing taxes on negative ones and subsidies (``bounties'') on positive ones.

Coase's critique of the Pigovian framing of environmental problems focuses on the nature of the transfer payment required to internalize an externality. He argues that, because of the symmetry of the problem, a tax on producers of a negative externality is not the only possible solution: 
\begin{quote}
    ``The traditional approach has tended to obscure the nature of the choice that has to be made. The question is commonly thought of as one in which A inflicts harm on B and what has to be decided is: how should we restrain A? But this is wrong. We are dealing with a problem of a reciprocal nature. To avoid the harm to B would inflict harm on A. The real question that has to be decided is: should A be allowed to harm B or should B be allowed to harm A?''
\end{quote}
That is, Coase takes issue with Pigou's premise that the one who causes the externality should be the one who is rewarded (if the externality is positive) or penalized (if the externality is negative). Judging from the textbooks reviewed above, Coase's criticism of Pigou's asymmetric treatment of pollutee and polluter is perhaps less well-known, but it was a key point in his seminal paper. To demonstrate the feasibility of alternative regimes, Coase discusses the different treatment under common law of escaped domesticated and wild animals. If a domesticated animal escapes and does damage, its owner is liable. If a wild animal escapes from captivity and does damage, the victim is liable rather than the former captor. Coase underlines the arbitrary nature of this distinction by discussing the rabbit, which many people would think of as \textit{domesticated} (and tame), but is actually a \textit{wild} animal under common law.

Coase' central example is cattle eating a neighbor's crops. Coase argues that, if the cattle-owner is liable for the damage done by her steers, she would limit the size of her herd to the point where the damage done by one additional steer equals the cattle's incremental profit. Coase then argues that, without such liability, the farmer would be willing to pay his neighbor to reduce the herd size, and that he would pay up to the point where the damage avoided by one fewer steer equals the marginal steer's value to the cattle-raiser. In other words, the final outcome is the same regardless of whether or not the cattle-owner has a duty to compensate for harm to her neighbor.

This example leads to the Coase Theorem: In the presence of externalities, clearly defined property rights, and the absence of transaction costs, agents can bargain their way to a Pareto optimum, and that Pareto optimum is the same regardless of who imposes an externality on whom.

Coase emphasizes that his conclusion only holds if there are no costs involved in the transaction and that it is easier to reach agreement if fewer parties are involved. He implicitly assumes that people are well-informed, act in their self-interest, that the money changing hands does not affect the demand or supply curves, and that the agreement reached by the bargaining parties will be enforced by courts if necessary. Another assumption is that the willingness to pay to avoid harm is equal to the willingness to accept compensation for harm. Coase himself did not seem to believe that these conditions were likely to be met in most situations, emphasizing the importance of considering the net value of alternative (imperfect) institutions that can be implemented in the presence of transaction costs. In his Nobel Prize lecture, he said that ``the legal system will have a profound effect on the working of the economic system and may in certain respects be said to control it'' \citep{Coase1991}.

In his seminal paper showing that competitive equilibrium with a Pigou tax is a Pareto optimum, \citet{Baumol1972} attacks Coase, writing
\begin{quote}
    ``It is ironic that just at the moment when the Pigovian tradition has some hope of acceptance in application it should find itself under a cloud in the theoretical literature. [...] Ronald Coase has asserted that the tradition has not selected the correct taxation principle for the elimination of externalities, and may not even have chosen the right individuals to tax or to subsidize.''
\end{quote}
Baumols' words may partly explain the divergent opinions of environmental economists on the Coase Theorem, as illustrated by how the theorem is treated by different environmental economics textbooks. However, Baumol does not address Coase' core contentions, which involve the initial allocation of property rights and the effect of transaction costs. Baumol is interested ``in the large numbers case'', where bargaining is impractical\textemdash whereas Coase is interested in externalities between a few agents. Correcting \citet{Buchanan1962}, Baumol shows that compensation of victims is not optimal \textit{at the margin} but adds, on page 312, ``except, of course, for lump sum payments.'' Baumol's objections are therefore not of a technical nature, but rather pragmatic: Coase caused confusion among policy makers who were just beginnning to accept Pigovian taxes. Indeed, Coase himself argued that ``[t]he significance to me of the Coase Theorem is that it undermines the Pigovian system'' \citep{Coase1991}.\footnote{\citet{Lai2011} and \citet{Medema2020} review how Coase' work was received by other economists, highlighting how different people emphasized different aspects of the Coase Theorem.}

Coase's Nobel Prize lecture continues: ``[s]ince standard economic theory assumes transaction costs to be zero, the Coase Theorem demonstrates that the Pigovian solutions are unnecessary in these circumstances. Of course, it does not imply, when transaction costs are positive, that government actions (such as government operation, regulation or taxation, including subsidies) could not produce a better result than relying on negotiations between individuals in the market.  Whether this would be so could be discovered not by studying imaginary governments but what real governments actually do. My conclusion; let us study the world of positive transaction costs.'' \citep{Coase1991}. That is, in his Nobel Lecture, Coase does not take issue with Pigou, but rather with the assumption of zero transaction costs. The somewhat arbitrary allocation of property rights is no longer central in Coase's mind.

\subsection{Coase formalized}
\citet{MasColell1995}, among others, state the Coase Theorem formally. Consider two agents with an indirect utility function
\begin{equation}
    v_i(p,w_i,h) = \max_{x_i \geq 0} u_i(x_i,h) \text{ s.t. } px_i \leq w_i \text{ for } i=1,2
\end{equation}
where $p$ is the price vector for consumption bundle $x_i$ of agent $i$, $w_i$ is his budget constraint, $u_i$ is utility, $v_i$ is indirect utility and $h$ is the externality. Assuming a quasilinear utility function with respect to a numeraire, we can write $v_i(p,w_i,h) = \phi_i(p,h)+w_i$. If both agents are price-takers, we can write $\phi_i(p,h)$ as simply $\phi_i(h)$.

Suppose that agent 1 chooses $h$ to maximize $\phi_1$. Then $\phi'_1(h^*) = 0$. The social optimum maximizes $\phi_1+\phi_2$, so that $\phi'_1(h^\circ) = -\phi'_2(h^\circ)$. The equilibrium $h^*$ is suboptimal unless $h^*=h^\circ=0$. If $\phi'_2(.)<0$, the externality is negative and $h^* > h^\circ$, that is, agent 1 chooses too much $h$. If $\phi'_2(.)>0$, the externality is positive and $h^* < h^\circ$, that is, agent 1 chooses too little $h$.

Now suppose that agent 2 has the right to be free of externality $h$, but would be prepared to waive that right in return for compensation $T>0$. Then agent 2 would solve
\begin{equation}
    \max_h \phi_2(h) + T \text{ s.t. } \phi_1(h) - T \geq \phi_1(0)
\end{equation}
where the constraint comes about because agent 1 needs to agree to the bargain. As the constraint binds, this is equivalent to
\begin{equation}
    \max_h \phi_2(h) + \phi_1(h) - \phi_1(0)
\end{equation}
The maximand is the social welfare function (shifted by a constant), and thus the equilibrium externality is the optimal one $h^\circ$.

If instead there are no restrictions on agent 1, agent 2 would need to compensate her with an amount $T<0$. Agent 1 would agree if $\phi_1(h)-T \geq \phi_1(h^*)$. Deciding on the offer made, agent 2 would solve
\begin{equation}
    \max_h \phi_2(h) + \phi_1(h) - \phi_1(h^*)
\end{equation}
The maximand is again the social welfare function (shifted by a different constant), and the equilibrium externality is the optimal one $h^\circ$.

This simple proof of the Coase Theorem also reveals key underlying assumptions:
\begin{enumerate}
    \item \textbf{No wealth effect} Quasi-linearity in the numeraire makes the externality $h$ independent of budgets $w_i$ and side-payment $T$.
    \item \textbf{Perfect information} The agents know each other's indirect utility functions.
    \item \textbf{Rationality} Agents maximize utility.
    \item \textbf{No endowment effect} The utility functions are smooth in the status quo, and economic agents behave the same whether or not they have the right to be free of externalities.
    \item \textbf{Zero transaction costs} The bargain can be struck without incurring costs.
\end{enumerate}

\citet{Medema2020} splits the Coase Theorem into three parts. The \textit{efficiency thesis} states that, once property rights are assigned, a Pareto optimum is achieved. As the assignment of property rights completes the market, this result is equivalent to the First Fundamental Theorem of Welfare Economics. The \textit{invariance thesis} states that the Pareto optimum is independent of the initial allocation, a result that is sharper than the Second Fundamental Theorem of Welfare Economics.

\textit{Zero transaction costs} is the third\textemdash and most controversial\textemdash part.\footnote{\citet{Lai2011} splits the Coase Theorem into two parts: efficiency and invariance.} Large parts of economic theory assume that transaction costs are negligible. If so, the Coase theorem illustrates that there is no need for direct government intervention to internalize externalities.

\subsection{Coase generalized}
\citet{Ellingsen2016} offer an alternative, formal proof of the Coase Theorem that is more general than the one by \citet{MasColell1995} shown above. \citet{Ellingsen2016} also show that the Coase Theorem only holds for two economic agents\textemdash one polluter and one pollutee, a $1\times1$ bargain. If there is more than one person involved on either side\textemdash $m\times1$, $1\times n$ or $m\times n$ bargains\textemdash then coordination problems between polluters or pollutees prevent the attainment of an efficient solution.

As a corollary, if there is no coordination problem, the Coase Theorem does hold for more than two agents. For instance, a $1\times n$ bargain between 1 polluter and $n$ pollutees is equivalent to $n$ $1\times 1$ bargains if there is no fixed cost of emission reduction, the variable costs are linear in emission reduction, the environmental damage is linear in emissions, and the polluter cannot exert market power over the polluttees. Under these (stringent and unrealistic) assumptions, each pollutee would strike a separate bargain with the polluter and those bargains would be efficient as the pollutees do not affect each other.

In more realistic settings, the action of one pollutee does affect the other pollutees\textemdash or polluters may affect each other. This would be the case if, for instance, the impact of pollution is non-linear in emissions. Then, coordination problems arise, and a pollutee may choose to free-ride on the efforts of her fellow pollutees to bargain with the polluter. 

Coordination problems have been thoroughly studied and are hard to solve.\footnote{See \citet{Ellingsen2016} for a succinct but excellent literature review.} That said, \citet{Ellingsen2016} show that, while $m\times n$ bargains do not attain efficiency, they can still improve welfare. In the examples discussed below, we focus on coordination and improvements in welfare resulting from bargains between two or more actors, rather than on Pareto optimality. These can be thought of as impure forms of the Coase Theorem, or examples of Coase-like bargaining that do not necessarily result in a Pareto optimum.

In the context of common pool resource management (oil exploitation), \citet{Libecap1984} show that 4-5 companies could negotiate an agreement and that 10-12 companies could only agree to request state intervention. \citet{Wiggins1985} add that asymmetric information hampers private contracting for oil exploitation in a common pool. \citet{Libecap1984} also find that, if there are more than 12 parties, no agreement at all could be reached. \citet{Libecap1985} argue that this is because oil companies are sufficiently influential to block state intervention.

\section{Coase in the lab}
\label{sc:lab}
Since the assumptions underlying the Coase Theorem were first made explicit, many laboratory experiments have been designed to understand which of these assumptions are mathematically convenient but can be relaxed without overturning the practical implications of the Coase Theorem and which assumptions are crucial. We briefly summarize their conclusions in this section, focusing on which conditions have proven to be particularly important for achieving Pareto optimal outcomes. 

Although numerous bargaining scenarios had been tested experimentally in the 1960's and 1970's, \citet{Hoffman1982} are generally credited to be the first to explicitly test, and confirm, the Coase Theorem in the lab.\footnote{\citet{PRUDENCIO1982} was published in the same year but seems to be older. Prudencio finds that a contract negotiated over an externality comes, on average, within 3\% of the Pareto optimum, and that there is no statistically significant difference between cases where the polluter or pollutee holds the initial property rights. Unfortunately, Prudencio's experiment ends with an ultimatum, and his players appear to be motivated by fairness as well as efficiency.} Subjects were assigned to groups of two or three. One or two subjects were randomly assigned to be ``controllers'', who, analogously to being assigned initial property rights in the Coase Theorem, had the right to unilaterally choose the set of payoffs players would receive. The other participant(s) could attempt to influence the outcome via negotiations, including by offering to transfer some or all of her earnings to the controller.\footnote{No physical threats were allowed.} In each case there was a unique scenario that maximized total cash payments, but whether or not payments were known to all participants varied. Any contract between the players was enforced by the experimenter, and payments were made publicly. Under conditions where payoffs were known and there was only one controller, 89.5\% of the 114 experimental decisions resulted in Pareto optimal outcomes. In experiments with limited information and joint controllers, success rates were substantially lower.

Since then, many experimental studies of the Coase Theorem and its limitations have been conducted, yielding much insight about when property rights are sufficient to yield Pareto optimal outcomes.\footnote{\citet{Harrison1985} refine Hoffman's experimental design to make cooperation individually rational. In the original set-up, it was impossible to distinguish between a fair allocation and a Pareto optimal one. They find strong support for the Coase Theorem: The Pareto optimum is found in 97\% of experiments.} Four review articles that summarize key experiments exist. \citet{Bohm2003} reviews seven experimental papers published between 1982 and 1998. For zero transaction costs, complete information, and small incentives, subjects tend to opt for a fair allocation rather than a Pareto optimal one. Higher incentives lead to a shift to the Pareto optimum. The Pareto optimum becomes unattainable if transaction costs increase.

\citet{Camerer2007} review these and later experiments. They report that private (rather than public) information does not affect the ability of participants to attain the Pareto optimum. Asymmetric information does: Participants are less willing to trade in this case. Less secure property rights attenuate the effect of asymmetric information. \citet{Camerer2007} also speculate that endowment effects would hamper Coasian bargaining.

\citet{Croson2009} reviews largely the same literature but with a different focus: She emphasizes that the Coase Theorem holds also when stress-tested with larger numbers of participants, asymmetric payoffs, uncertain payoffs, and more complicated bargaining. Finally, \citet{Prante2007} conduct a meta-analysis of experimental results, with the probability of obtaining the Pareto optimum as the dependent variable. They find that transaction costs and time-limits have a negative effect on that probability, while face-to-face bargaining and information have a positive effect. 

These four survey papers establish that, at least in the lab, the Coase Theorem holds under its original assumptions\textemdash and that it sometimes holds under conditions that are less strict. With fairly comprehensive results in the literature, it is no surprise that few recent papers have experimentally tested the Coase Theorem. We found one.\footnote{\citet{Galiani2014} study the Political Coase Theorem, with a focus on commitment. \citet{Hong2016} assess voluntary participation in public goods provision with Coasian bargaining.} \citet{Bar-Gill2016} find that the Coase Theorem also holds if either party can block the transaction and have the experimenter take away the good that they are bargaining over with mininal compensation. As above, a deviation from the strict assumptions of the Coase Theorem does not necessarily mean that its basic implications collapse.

\section{Coase and the courts}
\label{sc:legal_issues}

To understand whether the experimental results discussed above have empirical counterparts, it is worth considering whether the world's legal institutions are conducive to Coasian bargaining. While a full review of legal systems is beyond the scope of this paper, the US, which we focus on here, offers an illustrative and important example.

Well-defined property rights (and, implicitly, enforceable contracts) are the key assumption underlying Coasian bargaining. In legal systems with strong protection of private property, such as the United States, clearly defining property rights may seem straightforward. However, specifying complete property rights requires attention to such details as mineral rights, wildlife harvesting rights, rights to make noise or emit noxious smells, and so on. As court cases demonstrate, there are many situations in which property rights are sufficiently vague to result in substantial disagreements between the affected parties about who holds a particular right. For simplicity, we will refer to the party producing an environmental externality as the ``polluter'' and the party experiencing the environmental externality as the ``pollutee''. 

There are at least four reasons for the continued existence of ambiguous property rights. First, the common law theory of nuisance makes it very difficult to fully and clearly assign the right to create or assign the externality to the polluter, particularly for new types of harms where precedent has not been established. Second, it is difficult to define terms used in legislation and regulation in a way that leaves no room for an alternative interpretation. Third, the existence of multiple levels of government and of multiple, related, laws sometimes creates ambiguity about which law applies to a particular situation. Fourth, new laws and regulations change property rights, and shifting social norms and legal principles change what is deemed permissible.

The (very old) common law principle of nuisance is the basic legal principle determining the allocation of property rights around externalities from private property \textemdash who has the right to pollute and who has the right to be protected from pollution? In common law, the tort of nuisance goes back to the 13th century, in a case where King John of England (of \textit{Robin Hood} and \textit{Magna Carta} fame) ruled in favour of Simon of Merston after Jordan the Miller had flooded Simon's land in an attempt to expand the pond that powered Jordan's mill \citep{Brenner1974}. Since the resolution of the Trail Smelter dispute, in which the smoke of a lead and zinc smelter in British Columbia affected farmers in Washington, the legal obligation to be a good neighbour also applies across country borders \citep{Kuhn1938}.

In modern legal theory, the nuisance principle allows for the ``quiet enjoyment'' of private property, while protecting other people from  ``unreasonable interference'' as a result of that enjoyment. However, these are vague and general principles. In many situations, what constitutes ``unreasonable inteference'' is unclear, or at least contested, resulting in both polluters and pollutees asserting that they hold the right to inflict the nuisance or to be free from it, respectively \citep{Farber2019}. These cases sometimes lead to costly nuisance lawsuits, requiring a judge to weigh in to resolve the ambiguous allocation of rights. 

Similar issues surface in other environmental settings. There have been lawsuits over the exact definitions of ``discharge'' (S. D. Warren Co. v. Maine Board of Environmental Protection, 547 U.S. 370 (2006)), ``fill material'' (Coeur Alaska, Inc. v. Southeast Alaska Conservation Council, 557 U.S. 261 (2009)), ``navigable waterway'' (Rapanos v. United States, 547 U.S. 715 (2006)), ``flood or flood waters'' (Central Green Co. v. United States, 531 U.S. 425 (2001)), and ``acceptable noise.''\footnote{In the countryside, you should not complain about chickens clucking. See Anon, \href{https://www.lto.de/recht/nachrichten/n/lg-koblenz-6s21-19-nachbar-dorf-huehnerhaltung-laerm-zumutbar/}{Auf dem Land ist H\"{u}hner gackern \"{u}blich}, Legal Tribune Online, 10 Dec 2019.}

In other cases, there is sufficient ambiguity over which laws apply in a particular situation to prompt a costly lawsuit. In one case, ultimately decided by the US Supreme Court, the question at hand was whether the state or the Environmental Protection Agency determined what constitutes the ``best available control technology'' (Alaska Dept. of Environmental Conservation v. EPA, 540 U.S. 461 (2004)). The US Supreme Court has also ruled on whether a federal law governing pesticide law preempted farmers from suing under state law (Bates v. Dow Agrosciences LLC, 544 U.S. 431 (2005)). In another case, the US Supreme Court was asked to rule on whether the Endangered Species Act imposed additional requirements on activities governed by the Clean Water Act (National Assn. of Home Builders v. Defenders of Wildlife, 551 U.S. 644 (2007)). 

The considerations above result in imperfectly defined property rights and therefore inhibit Coasian bargaining, at least before precedent has been established through the courts. Determining whether these barriers can be resolved is beyond the scope of this paper.

Furthermore, property rights are also not immutable. Rewilding is one example. Large grazers were introduced in many nature reserves in Western Europe to keep landscapes open. Large predators are now being introduced to prevent overgrazing. As these wolves also kill the occasional sheep, the European Union now recommends full compensation for lost livestock.\footnote{See \href{https://ec.europa.eu/info/news/amendments-state-aid-guidelines-agriculture-sector-better-address-damages-caused-wolves-and-other-protected-animals-2018-nov-08_en}{CEC DG Agriculture and Rural Development, 8 Nov 2018}.} Rhineland-Palatinate guarantees compensation.\footnote{See Ministerium f\"{u}r Umwelt, Landwirtschaft, Ern\"{a}hrung, Weinbau and Forsten, 2015, \href{https://mueef.rlp.de/fileadmin/news_import/Wolfmanagmentplan.pdf}{Managementplan f\"{u}r den Umgang mit W\"{o}lfen in Rheinland-Pfalz}.}  A customary privilege of safety for farm animals has been replaced by an explicit right to compensation.

Environmental standards, in particular, are generally tightened over time. This trend implies that rights to pollute tend to disappear and rights to be free of pollution tend to appear. Loosened regulations have the opposite effect. When governments tighten environmental regulations, compensation may be offered to the companies newly deemed to be polluters. Recent examples include more stringent standards for nitrate emissions \footnote{See Anon, \href{https://rp-online.de/nrw/staedte/juechen/juechen-nrw-finanzminister-hilft-landwirten-die-sich-wegen-der-neuen-duengeordnung-um-ihre-existenz-sorgen_aid-47910943}{Landwirte sorgen sich um ihre Existenz}, Rheinishe Post Online, 28 Dec 2019.} and odour from farms,\footnote{See J. de Vries, \href{https://www.volkskrant.nl/nieuws-achtergrond/regering-gaat-varkenshouders-uitkopen-boeren-kritisch-over-plan~b925fcee/}{Regering gaat varkenshouders uitkopen, boeren kritisch over plan}, Volkskrant, 8 Jul 2018.} as well as pesticide bans,\footnote{See S. H\"{a}ne, \href{https://www.tagesanzeiger.ch/schweiz/standard/bund-verbietet-risikopestizid-bauern-verlangen-entschaedigung/story/25825599}{Bund verbietet Risiko-Pestizid – Bauern verlangen Entsch\"{a}digung}, Tagesanzeiger, 12 Dec 2019.} with politicians promising to make farmers whole. This is not Coasian bargaining\textemdash which is bargaining \textit{given} initial property rights\textemdash but rather bargaining over \textit{the assignment} of initial property rights\textemdash meta-Coase bargaining, if you will. There is little economic analysis of bargaining over the assignment of initial property rights \citep{Colby1995}.

Social norms can also play a role in defining the terms on which externalities are bargained over. Protests against and boycotts of large polluters have a long history \citep{Delacote2009, Olzak2009}. Like property rights, social norms can also shift over time with evolving standards of what constitutes a permissible nuisance as opposed to unacceptable behavior. Examples include shifting social norms around public littering or the disposal of dog waste. Decades ago, individuals had the ``right'' to dispose of waste in public spaces creating disamenities for others. But changing attitudes, sometimes driven by deliberate messaging campaigns and often codified in local laws and ordinances, shifted so that instead people now generally internalize at least some of the costs of responsible waste disposal while in public areas. As another example, the \textit{Stop the Child Murder} movement in the Netherlands ensured that road safety standards were enforced \citep{Reid2017}. Similarly, China's Center for Legal Assistance to Pollution Victims focuses on the enforcement of existing environmental legislation \citep{Xu2006}. Lawsuits against emitters of carbon dioxide seek to establish a legal right to an unchanging climate \citep{Tol2004, Peel2018}.

\section{Coase in the wild}
\label{sc:wild}
Strictly, the Coase Theorem applies to a bargain between two players who have no other interactions and do not expect to meet again. Such conditions can be approximated in the lab, but are rarely if ever met in reality. Furthermore, with the exception of countries negotiating over transported emissions, there are few interesting environmental problems with only two agents. As illustrated by experimental evidence, however, the strict requirements of the Coase Theorem can in some cases be relaxed without jeopardizing its applicability. We therefore also include examples that involve more than two agents.

In the Coase Theorem, polluter and pollutee are symmetric in the sense that the Pareto optimum will be reached regardless of how property rights are initially endowed. From the point of view of the parties involved, however, the endowment of initial property rights is critical in the sense that it determines who is imposing the externality on whom, and therefore the direction of the transfers involved in the Coasian bargain. Ronald Coase was keenly aware of this, as evidenced by his detailed discussion of the differential treatment in common law of harm caused by escaped domesticated animals and wild animals in captivity. There are also long-established legal and moral principles about harm and nuisance to third parties. These legal doctrines of nuisance delineate the property rights relevant to the negotiation of externalities between parties, though are often open to different interpretations, as described in the previous section. As there is a difference between the polluter and the pollutee paying, we discuss them separately, starting with the polluter.

\subsection{Polluter pays}
We first discuss examples of Coasian bargaining where the polluter ended up paying. Frequently, even if courts were not involved, the prospect of legal recourse to enforce property rights via a lawsuit lurked in the background. For example, in 2002, American Electric Power bought all 90 houses in Cheshire, Ohio, and all 221 residents left after health concerns were raised about the release of fly ash from the nearby coal-fired Gavin Power Plant \citep[][p. 262]{Kolstad2011}. Homeowners were compensated well above the market value. No lawsuit was filed, but the lawyers negotiating on behalf of the town did threaten to.\footnote{See K.Q. Seely, \href{https://www.nytimes.com/2002/05/13/us/utility-buys-town-it-choked-lock-stock-and-blue-plume.html}{Utility buys town}, New York Times, 13 May 2002.}

The American Electric Power company was certainly not the first to take this approach. Dow Chemicals, Georgia Gulf, Exxon, Shell, and Conoco have all bought properties near their chemical plants and refineries. Exxon and Shell appear to have started such purchases after explosions at their facilities caused damage to the people living nearby. Georgia Gulf's program began after a 1987 lawsuit, settled out of court, over contamination and health complaints. Reveilletown, Louisiana, no longer exists after Georgia Gulf bought it. Conoco's program is also in response to a lawsuit.\footnote{See K. Schneider \href{https://www.nytimes.com/1990/11/28/us/chemical-plants-buy-up-neighbors-for-safety-zone.html}{Chemical Plants Buy Up Neighbors for Safety Zone}, New York Times, 28 Nov 1990} Dow Chemicals' program was in response to \textit{the threat of} a lawsuit after chemicals spilled into the drinking water of Morrisonville, Louisiana. The town was abandoned in 1993. These are all examples of the polluter agreeing to pay the pollutee, under the threat that property rights established under the nuisance doctrine would be enforced in court. 

These are cases with one polluter and many pollutees. The coordination problem was solved by the polluter. The outcome need not be efficient, because the polluter had monopsony power.

Examples outside of the US include Severonickel, a copper-nickel smelter on the Kola Peninsula in Russia, which pays the nearby Lapland Biosphere Reserve \$300,000 annually, following a settlement in a court case Severonickel was likely to lose \citep{Shestakov2000}. Schiphol Airport in the Netherlands is planning to buy out homeowners troubled by the noise from an increase in the number of flights. The airport cannot grow without permission from the municipality of Aalsmeer, the local electorate is concerned about noise, and local politicians worry about re-election.\footnote{See Anon, \href{https://www.nhnieuws.nl/nieuws/230960/schiphol-en-aalsmeer-praten-over-uitkopen-bewoners-vanwege-geluidsoverlast}{Schiphol en Aalsmeer praten over uitkopen bewoners vanwege geluidsoverlast}, Noord-Holland Nieuws, 18 Sep 2018.} Similarly, the Government of Berlin financially compensates homeowners for the noise it permitted Tegel Airport to make.\footnote{See D. Bath, \href{https://www.morgenpost.de/berlin/article227941793/Tegel-Senat-gibt-3-5-Millionen-Euro-fuer-Laermschutz.html}{Tegel: Senat gibt 3,5 Millionen Euro f\"{u}r
L\"{a}rmschutz}, Berliner Morgenpost, 17 Dec 2019} The Royal Norwegian Air Force has bought houses near its {\O}rland base and paid for noise insulation for houses further afield.\footnote{See C. Ellingsen, N. Reinertsen and A.L. Kumano-Ensby, \href{https://www.nrk.no/dokumentar/xl/stoy-pa-landet-1.12898727}{St{\o}y p{\aa} landet}, Norsk rikskringkasting, 19 Apr 2016.} The US Air Force, by contrast, had to be ordered by the courts to pay compensation to the people living near the Yokota air base in the outskirt of Tokyo.\footnote{See Anon, \href{https://www.japantimes.co.jp/news/2017/10/11/national/crime-legal/government-ordered-pay-damages-aircraft-noise-u-s-yokota-base/\#.XnZiaYj7RPY}{Government ordered to pay damages over aircraft noise at U.S. Yokota base}, Japan Times, 11 Oct 2017.}

Chlorides in the Rhine river are another example of Coase-without-courts \citep{Bernauer1995, Phaneuf2017}. After concerns were raised about local groundwater contamination, \textit{Mines de Potasse d'Alsace} (MdPA) has, since 1931, dumped chlorides, a waste product of its potassium mining, in the Rhine instead, damaging farming and drinking water production downstream in the Netherlands. In the early 1970s, MdPA was the largest point source of chlorides, contributing 30-40\% of the load. Companies in Germany and Switzerland also dumped chlorides in the Rhine. In 1972, an agreement was reached between the governments of France, the owner of MdPA, the Netherlands, Germany and Switzerland to jointly compensate MdPA for the profits lost to emission reduction. France, the polluter, covered 30\% of the costs and the Netherlands, the pollutee, 34\%. Germany and Switzerland covered the remainder. These countries would rather pay MdPA to clean up its act than compel companies in their own countries to do the same. The 1972 agreement was revised in 1991. Switzerland now contributes less (3\% instead of 6\%) because a soda factory, its main source of chlorides, had closed. A quarter of the available funds was diverted from reducing pollution at its source in France to water purification in the Netherlands, as this had become economical since the original agreement. While the 1972 agreement mixed payments by polluters and pollutee, after 1991 the polluters paid almost all (91\%) of the costs of emission reduction. Transaction costs were high\textemdash at one point, the Netherlands recalled its ambassador to France\textemdash but not so high that it stopped negotiations. This example also highlights, as \citet{Coase1960} did, that mitigation is as important as compensation.

\citet[][p. 206]{Field2009} cite another example. The US Clean Water Act of 1972 empowered the Army Corps of Engineers to block development if that would damage wetlands. With the property rights firmly established, barter emerged. However, the Army Corps of Engineers cannot take money from developers. Instead, between 1993 and 2000, the Corps granted permits to damage some 24,000 acres of wetlands. In return, developers spent over \$1 billion to create, restore, improve, or protect about 42,000 acres of wetlands \citep{Bayon2004}.\footnote{This program has attracted the attention of scholars in ecology, engineering and law but not, as far as we know, in economics.} \citet[][pp. 255-6]{Berck2011} document that similar barter is common under the US Endangered Species Act, where the Fish and Wildlife Service allows for habitat swaps via mitigation banking.

\subsection{Pollutee pays}
Next, we review examples of situations where the pollutee pays the polluter to reduce the harmful activity. With a few exceptions that we discuss first, these cases involve governments or non-governmental organizations making the payments. However, the government payments are distinct from Pigovian subsidies in that they are lump-sum rather than per-unit payments. 

In 2016, apartment owners in a loft building in New York got together and paid \$11 million for the air rights next door, so that a developer could not build a building that would spoil the view; contributions were larger for owners of apartments on higher floors.\footnote{See J. Goodstein, \href{https://www.nytimes.com/2019/07/22/nyregion/manhattan-real-estate-views-air-rights.html}{How much is a view worth in Manhattan?}, New York Times, 12 July 2019.} Similarly, Mark Zuckerberg has bought out neighbors in Palo Alto, at a cost of \$43.8 million, to protect both his privacy and security \footnote{See \href{https://money.cnn.com/2016/05/25/technology/mark-zuckerberg-palo-alto-house/index.html}{Mark Zuckerberg to tear down and rebuild four houses surrounding his home}.}

Frequently cited as an example of \textit{payments for ecological services} \citep{Engel2008},\footnote{To the best of our knowledge, \citet[][p. 43]{Phaneuf2017} were the first to connect payments for ecological services to the Coase Theorem.} the Vittel case \citep{Perrot2006, Depres2008, Phaneuf2017} can also be interpreted as a manifestation of the Coase Theorem. Vittel, now part of Nestl\'{e}, sells mineral water. Run-off from farms near its spring meant that there was too much nitrate in the water. This risked Vittel's brand and its legal designation as ``mineral''. As farm run-off was below the legal limit and land-zoning prevented the conversion of agricultural land to other purposes, Vittel bought out some farms and negotiated individual long-term contracts with 26 farmers; some farmers did not contract. Vittel made an upfront payment to the farmers, pays them an annual fee, and subsidizes labor and technical advice; contracted farmers can graze their animals on Vittel lands. In return, the farmers minimize the application of nitrogenous fertilizers. Nestl\'{e} has used a similar approach to protect its other brands.\footnote{See \href{https://www.nestle-waters.com/newsroom/news/agrivair-protecting-water-sources}{Agrivair} for a high-level overview.} Transaction costs are small relative to the value of branded water, and fell as Nestl\'{e} gained experience in bargaining. This is an example in which a single pollutee pays multiple polluters. The coordination problem between polluters was solved by the pollutee, not necessarily efficiently as the pollutee may have exercised monopoly power. An effort to unionize the farmers failed, because some farmers preferred acting independently \citep{Wiggins1985}. The bargaining power of farmers fell as other farmers contracted \citep{Depres2008}.

New York City followed an approach similar to Vittel's to protect the watershed supplying the City's drinking water \citep[][p. 60]{Harris2018}. By 2010, its Watershed Land Acquisition Program had purchased or obtained conservation easements on 100,000 acres (10\%) in the Catskill-Delaware watershed from which New York City draws 90\% of its drinking water.\footnote{See \href{https://www1.nyc.gov/site/dep/environment/extended-nyc-watershed-land-acquisition-program.page}{Extended NYC LAP}.} The program continues.\footnote{See \href{https://www1.nyc.gov/site/dep/environment/about-the-watershed.page}{About the watershed}.} The problem had arisen because Delaware County could meet new federal standards on drinking water and New York City could not. Purchasing land and changing its management to preserve drinking water quality, while expensive, was cheaper than building new water treatment plants \citep{Church2009}.

Japan's Green Aid Plan is another example of the pollutee paying to reduce emissions. Japan invested over \$500 million in energy efficiency and clean coal projects in seven other countries in Asia, which China receiving more than two-thirds of the total \citep{Evans1999}. Concerned about winds blowing sulphur emitted in China to Japan, the Cleaner Coal Program stimulates the adoption of desulphurization technologies in coal-fired power plants.\footnote{Note that Japan also funds environmental projects that do not directly benefit Japan \citep{Potter1994}.} The program covers training and technical assistance as well as the donation of equipment. The projects in the program met their objectives. Desulphurization techniques were not taken up by power plants outside the program \citep{Oshita2002}, suggesting it was Japanese funding rather than Chinese concerns about air pollution that caused the installation of scrubbers.

The Baltic Sea Action Plan is similar, but smaller \citep{Backer2010}. Funded by Sweden (\euro{}9 million) and Finland (\euro{}2 million), the program provides financial and technical support, particularly to reduce the discharge of nutrients into the Baltic Sea by Estonia, Russia, and several other countries.\footnote{See \href{https://www.nib.int/loans/loan_products/trust_funds/bsap_fund}{Baltic Sea Action Plan}.} Earlier, Sweden funded similar projects, not just for water but also for air pollution \citep{Lofstedt1995, Hassler2002}.

Not all attempts to pay for pollution reduction are successful. A decade-long attempt by Finland, supported by Norway and Sweden, to clean up sulphur emissions from iron mining and smelting in Karelia and nickel smelters on the Kola Peninsula, came to nothing \citep{Kotov1996}, partly because of the chaotic situation in post-Soviet Russia \citep{Darst2001} and partly because of the difficulty in writing and enforcing contracts in Putin's Russia \citep{Victor1999}. This example goes to the heart of Coase Theorem: Well-defined property rights and the enforcement of any resulting agreements are key to success.

It is often argued that the Coase Theorem only works with a small number of players: \citet{Coase1960} used two agents, as does every textbook exposition. If there are many pollutees, they would free-ride on buying out the polluter. The City Council of Santa Maria, California, circumvented this problem by imposing a tax on residents near a feedlot causing pungent smells, and using the revenue to pay the owner to cease operations \citep[][p. 259]{Kolstad2011}. The coordination problem between pollutees was solved by the local government.

In another example, the Nature Conservancy and Environmental Defense Fund, both non-governmental organizations, acted on behalf of many people worried about destructive bottom trawling for fish and shellfish and bought up fishing permits and harmful fishing equipment \citep[][p. 265]{Kolstad2011}.\footnote{See J. Christensen, \href{https://www.nytimes.com/2006/08/08/science/earth/08fish.html}{Unlikely Partners Create Plan to Save Ocean Habitat Along With Fishing}, New York Times, 8 Aug 2006.} An NGO in the Netherlands has been doing similar things since 1905, using donations to buy land to turn it into a nature reserve; the NGO now maintains almost 2.5\% of the country's area.\footnote{See Natuurmonumenten, \href{https://res.cloudinary.com/natuurmonumenten/raw/upload/v1558342716/2019-05/NM190106\%20JV\%202018_V8.pdf}{Annual Report 2018}.} The Nature Conservancy used a reverse auction to pay 33 rice farmers in California's Central Valley to flood 10,000 acres during February and March, a time crucial for migrating birds. The program cost less than \$100 per acre, a fraction of the costs of permanent wetland creation \citep{Hallstein2014}. In all of these cases, there are multiple polluters and multiple pollutees. An NGO put itself in between, a visible hand coordinating the Coase-like bargain. This is not likely to be efficient\textemdash the NGO has both monopoly and monopsony power and may well have motives other than the efficient coordination of bargaining. Nonetheless, all parties engaged voluntarily so the transaction are Pareto improving.

\section{Discussion and conclusion}
\label{sc:conclude}
The Coase Theorem comes in three parts. (\textit{i}) The \textit{efficiency thesis} extends the First Fundamental Welfare Theorem to cases where there are externalities: If property rights on an externality are clearly assigned, bargaining leads to a Pareto optimum. (\textit{ii}) The \textit{invariance thesis} sharpens the Second Fundamental Welfare Theorem. Regardless of whom these property rights are assigned to, the \textit{same} Pareto optimum is reached through bargaining. (\textit{iii}) The first two parts hold only for zero transaction costs, as Coase emphasized, and a number of other restrictive assumptions enumerated by subsequent researchers.

The practical implications of the Coase Theorem for Pigovian taxation are unclear. On the one hand, the Coase Theorem can be used to argue against Pigovian taxes\textemdash the government should assign property rights but not otherwise intervene. On the other hand, the Coase Theorem can also be used to argue for Pigovian taxes because transaction costs are positive. We see Coasian bargaining and Pigovian taxation as complements, not substitutes. If there are a large number of polluters and pollutees, Coasian bargaining is impractical; Pigovian taxes are not. However, Pigovian taxes can be impractical too, as when managing an idiosyncratic externality between neighbours, for example. At the extreme, Pigovian taxes can be impossible if there is no higher state authority able to impose taxes, as in the case of an externality between two neighbouring countries.

We survey the application of the Coase Theorem to environmental and resource management issues. The Coase Theorem means different things to different people. Strictly, the Coase Theorem states that an externality can be bargained away, and that, regardless of the initial allocation of property rights, the same Pareto optimum will be reached. Coase presented this argument as a viable alternative to Pigou taxation but only for a small number of actors. Coase originally emphasized the arbitrary nature of the initial allocation, but later argued that the Coase Theorem is a \textit{reductio ad absurdum} to show that transaction costs are key.

Textbooks in environmental economics reflect this ambiguity. Some authors present the Coase Theorem as a theoretical curiosity, others as a viable alternative to Pigou taxes. Most books, however, take a middle position, emphasizing the many and strict conditions under which the Coase Theorem holds.

Laboratory experiments show that these conditions are not nearly as strict in practice as they are in theory. The Pareto optimum is likely to be found if payoffs are uncertain or asymmetric, bargaining complicated or involve multiple people, and information private\textemdash but not if information is asymmetric or transaction costs large.

We document a number of real-world examples of applications of the Coase Theorem. Cases in which the polluter pays are hard to interpret. With one exception, the polluter appears to be paying to avoid a court order\textemdash but this just reflects that, in Common Law, since 1201, there is a well-established duty to not harm a neighbour under the nuisance doctrine. We also document cases in which the pollutee pays. These tend to be cases falling outside the nuisance doctrine for a number of reasons\textemdash either they are international, so outside of domestic property rights regimes, or the externality involved would not be considered a nuisance under current legal interpretation (e.g. conversion of bird habitat for agriculture decades ago, or small amounts of nitrate runoff).

One frequent objection raised to the idea of applying Coase as a substitute to Pigovian taxation is the issue of transaction costs. For many environmental problems, harms are diffused over many people, so that negotiating individual contracts to reach the Pareto optimum would be impractical and costly. In practice, our review reveals that most substantive applications of Coase-like bargaining involve an entity acting on behalf of the aggregated interests of a large population, substantially reducing the transaction costs involved and solving the coordination problem between agents. These entities fall into two main categories\textemdash governments acting as agents of their people (for instance in international examples or the New York City watershed case), and environmental groups acting on behalf of their members.

One caveat of our analysis is that we do not know how many Coasian bargains would be struck in a world with more or less explicit rules about compensation. In the United States and many other developed countries, laws codify not just property rights but also how much compensation must be paid if those property rights are violated.

The central insight of Coase is that people, organizations, or even countries can bargain over externalities to produce Pareto improvements if property rights are well defined, contracts are enforced, and transaction costs are relatively small. This theoretical result has been borne out in laboratory studies, showing that some of its strict conditions can be relaxed without materially affecting its key features. The practical importance of the result for addressing environmental problems is less clear. Although our review uncovered several cases of Coasian bargains being struck ``in the wild,'' these generally appear limited in application, except in the international context, and are generally a complement to, rather than a substitute for, other forms of environmental regulation.

\begin{notes}[Acknowledgements]
The authors would like to thank Soren Anderson for pointing us to Cheshire case, Michael Springborn for the Central Valley example, and to Ben Hansen for the brownies. Matthew Kahn was instrumental in the initial stages of this paper. Two anonymous referees had excellent comments that helped improve the paper.
\end{notes}

\bibliography{coase}

\appendix
\section{Coase in the classroom}
\citet{Hoyt1999} designed an experiment to demonstrate the Coase Theorem in the classroom. Students are split into two groups, and paired. One half of students is given pencil and paper, and rewarded for solving mathematical problems. The other half are given eraser and the paper used by their counterpart in the first group, and rewarded for folding paper planes \textit{without writing on}. After the completion of the first round, results are discussed and a collective decision is made who should compensate whom and by how much. In the second round, students can transfer part of their rewards to their counterparts, but the rules are otherwise unchanged.

\citet{Andrews2002} introduces two variants of \citet{Hoyt1999}'s experiment. Firstly, students in the first group have to buy either an expensive pencil or a cheap pen. This introduces a second mitigation policy (erasers are the first), giving more agency to the problem-solvers. Secondly, and more importantly, students are not paired. All paper used and unused by the first group is gathered up and students of the second group will have to scramble from the common pool. This makes the assignment of property rights harder. It moves the experiment away from the Coase Theorem, but arguably closer to reality.

\end{document}